# Line-by-line control of 10-THz-frequency-spacing Raman sidebands

**Kanaka Raju Pandiri[1], Takayuki Suzuki[1], Akira Suda[2], Katsumi Midorikawa[2], and Masayuki Katsuragawa[1, 3*]**

[1]*Department of Applied Physics and Chemistry, University of Electro-Communications, 1-5-1 Chofugaoka, Chofu-Shi, Tokyo, Japan 182-8585*
[2]*RIKEN Advanced Science Institute, 2-1 Hirosawa, Wako-Shi, Saitama 351-0198, Japan*
[3]*PRESTO, Japan Science and Technology Agency, 4-1-8 Honcho, Kawaguchi, Saitama, Japan*
[*]*corresponding author:* katsura@pc.uec.ac.jp

**Abstract:** We report line-by-line control of a coherent discrete spectrum (Raman sidebands) with a frequency spacing of 10.6 THz that is produced by an adiabatic Raman process. We show that the spectral phase of the Raman sidebands is finely controlled to the target (flat relative-spectral-phase). This is achieved by employing a combination of a spatial phase controller and a spectral interferometer, which are specifically designed for a high-power discrete spectrum. We also show that such spectral-phase control produces a train of Fourier transform limited pulses with an ultrahigh repetition rate of 10.6 THz in the time domain.

---

## 1. Introduction

There is increasing interest in synthesizing user-specified pulsed waveforms by manipulating phase and/or intensity in the frequency domain in accordance with Fourier transform relationships. Such optical pulse shaping has opened new possibilities [1, 2], such as pulse repetition-rate multiplication, arbitrary optical wave generation, and chemical reaction control. Optical pulse shaping has so far addressed spectral lines in groups, and recently pulse shaping under a new regime, line-by-line spectral manipulation, was demonstrated in several groups [3–6]. The most symbolic work is the processing of more than 100 spectral lines, reported by the Weiner group at Purdue University [6].

A coherent discrete spectrum produced in an adiabatic Raman process (i.e. a "Raman sideband") [7-9] is a good source for implementing such spectral line-by-line manipulation, since it has a discrete nature beyond the terahertz scale and has good mutual coherence between the spectral lines. Generation of a train of single-cycle pulses [10] and also its carrier-envelope phase control [11, 12] were recently demonstrated by synthesizing vibrational Raman sidebands with a liquid-crystal-based spatial light modulator. Here, we show the line-by-line control of rotational Raman sidebands with a frequency spacing of 10.6 THz [13, 14]. The advantage in this work over previous studies is our direct measurement of the discrete spectral phase on the basis of the concept of spectral interferometry, and the line-by-line control using a fused silica-glass-plate–based spatial phase controller. Estimation of the spectral phase is accurate and fast. Furthermore, the system withstands high-power laser (more than 1 J/cm$^2$) and short wavelength (near-vacuum ultraviolet). The synthesized waveforms will be applicable to high-energy-laser physics such as higher-harmonic generation.

## 2. Conceptual idea of line-by-line control of Raman sidebands

First, we describe the concept behind this study: how we generate Raman sidebands and then manipulate their spectral phase line-by-line. The conceptual layout of the whole system is illustrated in Fig. 1. It consists of two major parts: a system for Raman sideband generation and one for phase manipulation.

The generation of the Raman sidebands is based on adiabatic driving of the Raman coherence. The basic scheme is shown in the inset to Fig. 1. States $|g\rangle$ and $|u\rangle$ comprise a Raman-transition-allowed two-level system. We employ a pure rotational transition of J = 2 ← 0 (10.6236 THz) in parahydrogen for this Raman system. Then, we apply two pump-pulsed fields, $\Omega_0$ and $\Omega_{-1}$, between which the frequency difference is slightly detuned (+ 800 MHz) to the Raman resonance. This two-photon detuning is provided so that the adiabatic condition is satisfied with the smallest detuning—the optimal adiabatic condition [7, 13, 14]. When a sufficiently high intensity of pump field (typically, ~ $GW/cm^2$) is applied under this optimal adiabatic condition, a near-maximum Raman coherence can be created [7-9].

Once such a highly coherent molecular ensemble is realized, it acts as an optical modulator, similar to an electro-optic modulator. Namely, the molecular ensemble deeply modulates the transmitting light (i.e. in this scheme, the pump fields themselves). This results in the generation of discrete Raman sidebands. The frequency spacings of these sidebands are precisely equal (10.6228 THz), making a comb-like spectrum [15]. This nonlinear optical process is physically equivalent to that called "parametric stimulated Raman scattering." High coherence is what makes this system different from conventional parametric stimulated Raman scattering: with near-maximum coherence, the sidebands can be efficiently generated within a unit-phase slip length, enabling "coaxial generation of the Raman sidebands without any restriction by the phasematching condition" [7-9].

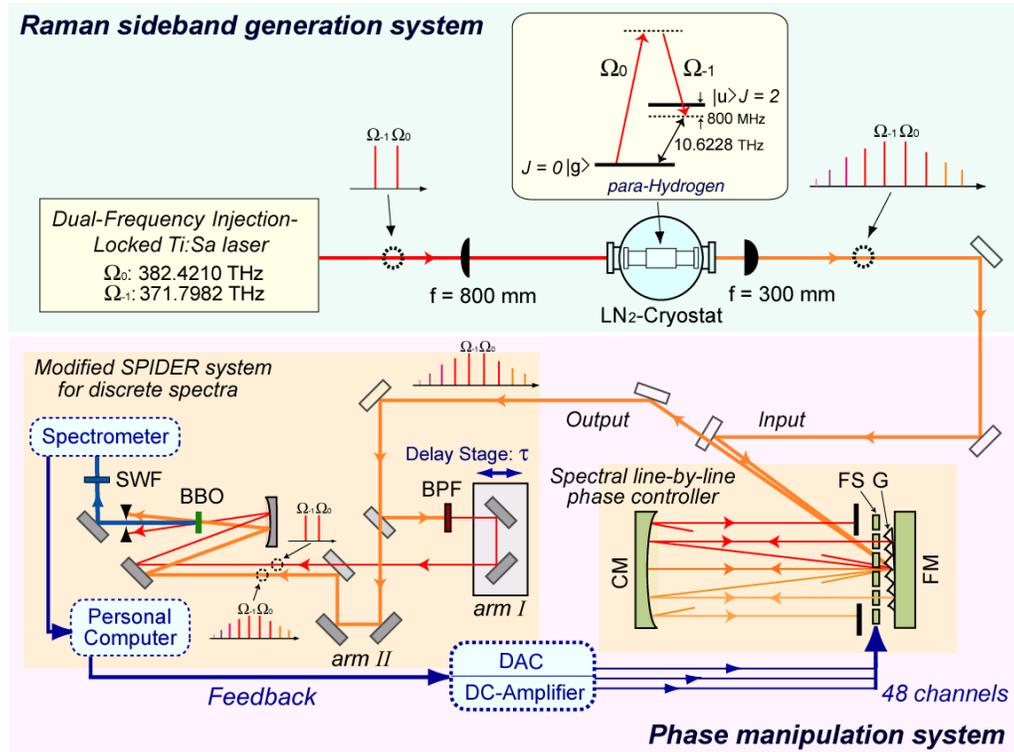

Fig. 1. Schematic of the experimental setup. $LN_2$-Cryostat: liquid-nitrogen cryostat; FM: folding mirror; G: grating; FS: fused silica plates; CM: concave mirror; Modified SPIDER system: modified spectral interferometry for direct electric-field reconstruction system; BPF: bandpass filter; BBO: β-barium-borate crystal; SWF: shortwave pass filter; DAC: digital-to-analog converter.

The generated Raman sidebands are mutually phase coherent in nature. Thus, their spectral phases can be controlled. Here, we manipulate them line-by-line. The brief setup is shown in Fig. 1 and is named the "spectral line-by-line phase controller" (SLLPC) [16]. Here, the Raman sidebands are resolved in space to the spectral lines. They are then focused to the plane (i.e. the Fourier plane) and their spectral phases are controlled individually by changing the optical-path lengths in the Fourier plane. Finally, they are combined again into a single beam in space.

The spectral phases of the Raman sidebands are monitored and estimated by using a spectral interferometry method. The system is called "spectral interferometry for direct electric field reconstruction for discrete spectrum (SPIDER-DS)". The basic principle is common to that of the original SPIDER [17]; the idea is, however, extended to estimate the spectral phase of a discrete spectrum [18, 19]. The point is to generate a pair of discrete spectra that are correlated with the original discrete spectrum and have an exact relative-frequency shift as their frequency spacing. Here, such a pair of spectra is produced by mixing the Raman sidebands with the pump fields $\Omega_{-1}$ and $\Omega_0$. Sum-frequency spectra having the same frequency spacing as that of the Raman sidebands, $\Delta\Omega = \Omega_0 - \Omega_{-1}$, are generated with an exact relative-frequency shift of $\Delta\Omega$. They interfere with each other and can provide a definite interference spectrum irrespective of randomness in the phases of the pump fields $\Omega_{-1}$ and $\Omega_0$. The interference spectrum is expressed [18, 19] as

$$I_n^{SF} = |E_{n+1}E_{-1}|^2 + |E_n E_0|^2 + 2|E_{n+1}E_{-1}E_n E_0|\cos(\Delta\Omega\tau - \Delta\varphi_n - \delta). \quad (1)$$

$E_n = |E_n|e^{i\varphi_n}$ is the complex amplitude of the nth order Raman sideband with a phase of $\varphi_n$. $\tau$ is the delay time applying to the pump fields, $\Omega_{-1}$ and $\Omega_0$, on one arm (arm I) of the interferometer (see Fig. 1). $\delta$ is the additional phase term which arises in the interferometer: it arises since the phase-difference term, $\varphi_0 - \varphi_{-1}$, is different between the arms I and II. The objective intrinsic phase-difference between the successive sidebands is given by $\Delta\varphi_n$ in this equation, which is defined by

$$\Delta\varphi_n = (\varphi_{n+1} - \varphi_n) - (\varphi_0 - \varphi_{-1}). \quad (2)$$

This $\Delta\varphi_n$ in Eq. (2) can be extracted by tracing the $I_n^{SF}$ as a function of "$\varphi_{n+1}$" or "$\tau$" in Eq. (1) (for more details see Ref. 18, 19).

Our final goal in this study was to control the relative spectral phases, $\Delta\varphi_n$, of the Raman sidebands to the target. The difference between the target and the measured relative spectral phase is detected by scanning the sideband phase, $\varphi_{n+1}$, over $2\pi$ and feeding the result back to the SLLPC as an error signal for the sideband, $\Omega_{n+1}$. The same process is consecutively applied to all the sideband components. The relative spectral phases, $\Delta\varphi_n$, controlled in this way are then estimated by scanning "$\tau$" in turn, as in the original SPIDER-DS mode [18, 19]. This series of processes is generally iterated until the measured spectral phases sufficiently coincide with the target ones. Depending on our purpose, we can set an appropriate target spectral phase. Here, we chose **"*flat relative spectral phase*"** (non-dispersive: $\Delta\varphi_n$ is zero for all of n) as a test target. This target implies "a train of Fourier-transform limited pulses with a time interval of the inverse of the frequency spacing of the Raman sidebands" in the time domain.

## 3. Experimental

In this section, we describe the details of the experimental system and also the procedure used for the actual experiment.

### *3.1 Detailed configurations of experimental system*

The pump laser for adiabatically driving the Raman coherence is a dual-frequency injection-locked laser [20-22]. It is composed of a power oscillator and two master lasers (external-cavity-controlled laser diodes). This pump laser simultaneously emits Fourier-transform-limited nanosecond pulses (typically 6 ns) at the two frequencies ($\Omega_{-1}$, $\Omega_0$). The carrier frequencies at the pulsed outputs can be continuously tuned with a precision better than 1 MHz by simply controlling the oscillation frequencies of the two master lasers. The key advantage of applying this injection-locked laser is that the temporal and spatial overlaps of the pulsed outputs at the two frequencies are satisfied perfectly, because they are generated from a single laser resonator.

The SLLPC is configured in accordance with a well-known 4f layout but folded at the Fourier plane [16]. The feature of this system is that 48 fused silica-glass plates are placed on the Fourier plane, instead of the usual liquid crystal array. Thereby, the system is applicable to a high-power laser, as demonstrated. The glass plates are 2 mm wide and 1 mm thick and are installed with a separation of ~0.2 mm from each other. The incident discrete spectrum (Raman sidebands) is dispersed with a grating (600 grooves/mm), and then each single-frequency component is focused onto the respective one glass plate with a concave mirror (f = 0.5 m). The spectral phases are manipulated by changing the angles of the glass plates (i.e., by changing the effective optical thickness) individually by using bimorph-type piezoelectric transducers (PETs). The PETs are driven by a set of 48-channel digital-to-analog converters (DACs) and amplifiers (applicable voltage range: –60 to 60 V) connected to a personal computer (PC). The folding mirror placed just behind the glass plates reflects the focused beams such that they return on the same optical path (slightly upward) as the incident light and are combined again into a single beam in space.

The SPIDER-DS has a configuration very similar to that of a standard autocorrelator. The difference is that the two frequency components ($\Omega_{-1}$ and $\Omega_0$) are extracted from the incident discrete spectrum on one of the two arms of the interferometer (arm I) [18, 19]. For this purpose, a bandpass filter is set on arm I. Both of the selected frequency components and the spectrum to be measured (Raman sidebands) on the other arm (arm II) of the interferometer are focused on to a 20-μm-thick β-barium-borate (BBO) crystal (Type I) with a concave mirror (r = 200 mm). The interference spectrum ($I_n^{SF}$) resulting from the pair of sum-frequency spectra generated in the BBO is measured with a compact spectrometer. An iris and a shortwave pass filter (SWF) are placed to block unnecessary frequency components. Finally, the measured interference spectrum is sent to the PC, which produces an error signal or estimates the spectral phase by using Labview-based software. The delay time τ is scanned by translating the corner mirrors on arm I, which are mounted on a single-axis precision stage.

*3.2 Operation of the experimental system*

The Raman medium, gaseous parahydrogen, was produced by converting normal hydrogen using a catalyst (oxidized-iron-powder) [23]. The purity of the parahydrogen was estimated to be greater than 99.9%. The parahydrogen produced was then confined to a cell (interaction length: 15 cm) at a density of $3 \times 10^{20}$ cm$^{-3}$. The parahydrogen was further cooled down to liquid nitrogen temperature to have a sufficient population probability of 97% in the ground state, $|g\rangle$.

Next, we set the oscillation frequencies of the dual-frequency injection-locked laser at 382.4210 THz (783.9331 nm) and 371.7982 THz (806.3312 nm); this provided an optimum two-photon detuning of + 800 MHz. These two pump laser radiations were then focused with the f = 800 mm lens into the parahydrogen. The peak intensity in the interaction region was set at ~4 GW/cm$^2$ (pulse energy: 3.8 mJ in total).

**4. Results**

Figure 2 (a) shows the generated Raman sideband spectrum. Twelve sidebands, from $\Omega_{-5}$ (329.3070 THz: 910.3738 nm) to $\Omega_6$ (446.1578 THz: 671.9426 nm), were observed. All the sidebands were generated coaxially with a high-quality beam profile (typical M$^2$ < 1.1 [13]). The generated sidebands were collimated with the f = 300 mm lens and guided to the phase-manipulation system. First, from among these 12, we picked up seven from $\Omega_{-4}$ (339.9298 THz: 881.9246 nm) to $\Omega_2$ (403.6666 THz: 742.6734 nm) by partly blocking the others in space in the SLLPC (see Fig. 1). The selected sidebands are highlighted by the dotted box. This selection was done to restrict the Raman sidebands to those that could produce sufficient SPIDER-DS signals. Figure 2 (b) shows the typical interfered sum-frequency spectra obtained in the SPIDER-DS. Each frequency component $I_n^{SF}$ was produced as a result of interference between the sum-frequency of $\Omega_n$ and $\Omega_0$ and that of $\Omega_{n+1}$ and $\Omega_{-1}$, as found in Eq. (1). Because seven sidebands were selected, six interfered sum-frequency components, $I_{-4}^{SF}$ to

$I_1^{SF}$, were obtained. The green dotted line shows those obtained at the initial stage of the control process.

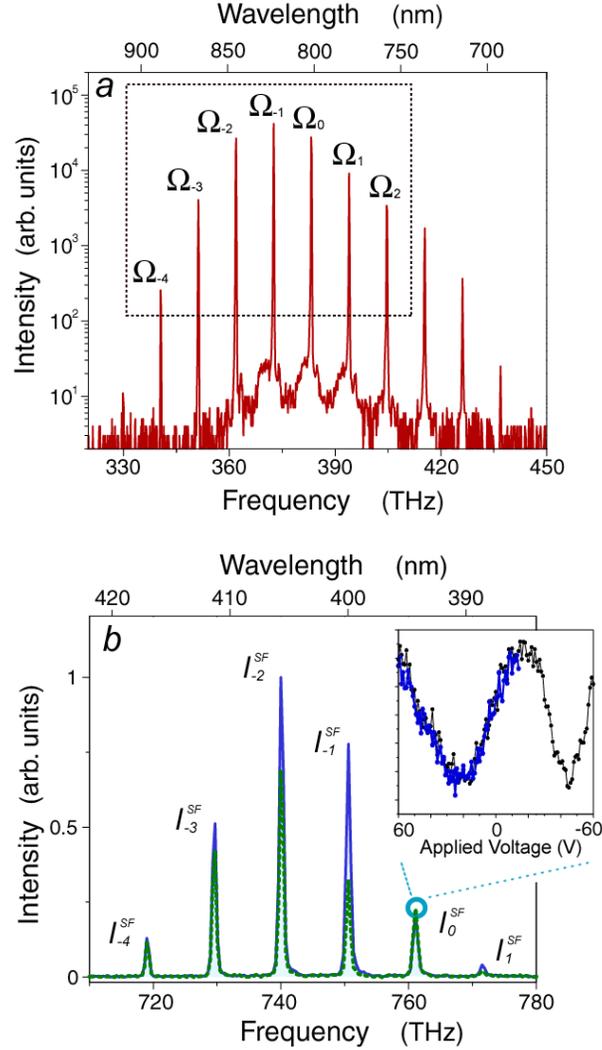

Fig. 2. a, Generated Raman sideband spectrum. Seven sidebands ($\Omega_{-4}$ to $\Omega_2$) were chosen for control of their spectral phases to the target. b, Interfered sum-frequency spectra obtained in the phase manipulation system. Each component is identified by the symbol $I_n^{SF}$ for n = –4 to 1. Green and blue lines are those before and after the control of the spectral phase to the target, respectively. $I_0^{SF}$ is zoomed out to show a typical peak intensity variation (black dots) as a function of voltage applied to the PET.

The actual control of the spectral phase to the target "flat relative spectral phase" was processed as follows. First, we set the delay stage (τ) at a position such that $I_{-1}^{SF}$ was at a maximum. This implies that we make the target relative spectral phase, "$\Delta\varphi_n = 0$," correspond to the "maximum" of $I_n^{SF}$ for all of n. This procedure was done to make the phase control more accurate in the actual process; it does not give any qualitative effect to the phase control. Then we manipulated each spectral phase to the target (maximum of $I_n^{SF}$) by changing the angle of the corresponding glass plate. One of such processes is shown for $I_n^{SF}$ in the inset to Fig. 2(b). The intensity (black line) varied sinusoidally with the voltage applied to the PET (–60 to 60 V). To reduce the intensity fluctuation, we averaged the measurement intensities over five shots. The target spectral phase (peak of the intensity) was quickly identified, and the corresponding voltage giving the target spectral phase was successfully applied back, as shown by the blue line. The voltage was applied unidirectionally to minimize the inherent hysteresis effect of the PET. The same procedure was followed with the other five sidebands. The interfered sum-frequency spectrum after control of the spectral phases to the target is represented by the blue solid line. It can be seen that all $I_0^{SF}$ were maximized (corresponding to the target relative spectral phase). The total time taken to complete this scheme was less than 10 min.

## 5. Estimation of spectral phase

After the above control process of the spectral phase to the target, we estimated the resultant relative spectral phase for the seven Raman sidebands. We used the same SPIDER-DS system employed in the control process, but it was operated with the original SPIDER-DS mode (scanning the time delay τ), as described in Sec. 2. Figure 3 shows the results of (a) before and (b) after control to the target. The measurement was done for 4.5 periodic cycles as a function of the delay time τ, whereas that for three cycles is shown here. The solid black lines are the fitting curves to the observed data (gray circles) with a sinusoidal function. Green and blue arrows indicate the minimum positions of the fitting sinusoidal functions for each of $I_n^{SF}$ components. It is clearly seen that the blue arrows in Fig. 3 (b) are nearly in the same position after control of the spectral phase to the target. This indicates that all the relative spectral phase, $\Delta\varphi_n = 0$. More correctly, the small variations in the relative spectral phase against the target are residual. This was mainly due to the precision of the phase control (~ ±0.1 rad), which was limited by the minimum step voltage (1 V) applied to the PET in the spectral line-by-line phase controller system.

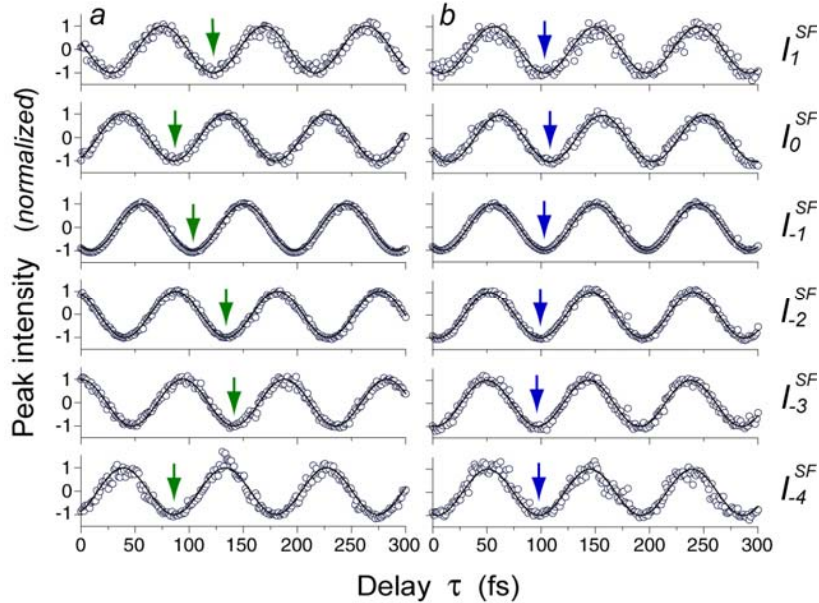

Fig. 3. Observed intensity variation of $I_n^{SF}$ with respect to the delay τ (gray circles) and the fitting curves with a sinusoidal function (black lines): (a) before and (b) after control to the target flat spectral phase. Green and blue arrows indicate minima of the estimated fitting curves.

On the basis of the estimated spectral phases in Fig. 3, we reconstructed the intensity waveforms consisting of the seven Raman sidebands in the time domain. The green and blue solid lines in Fig. 4 respectively represent the waveforms before and after control of the spectral phase to the target. We see that as a result of the control to the target relative spectral phase, a train of short pulses was formed with a 94-fs time interval (10.6-THz repetition rate), which precisely corresponded to the inverse of the frequency spacing of the Raman sidebands. The pulse duration was estimated to be 17 fs. The dotted line indicates the waveform calculated from the observed Raman sideband spectrum by assuming the Fourier-transform-limited condition (the target). The reconstructed pulse shape is in good agreement with the Fourier-transform-limited pulse shape, except for small deviations around the peaks and tails.

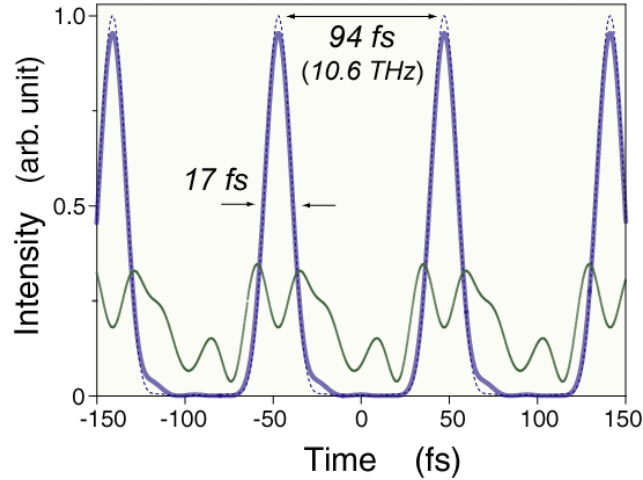

Fig. 4. Temporal intensity waveforms reconstructed on the basis of the estimated spectral phases of the Raman sidebands. Green and blue solid lines represent before and after optimization, respectively. Dotted lines show the Fourier-transform-limited waveform calculated from the measured sideband spectrum.

## 6. Summary

We showed line-by-line control of the Raman sidebands with a frequency spacing of 10.6 THz, generated by adiabatic driving of the rotational Raman coherence ($J = 2 \leftarrow 0$) in parahydrogen. We demonstrated that the spectral phase of the Raman sidebands were finely controlled to the target "flat relative-spectral-phase". The key techniques employed were the use of a spatial phase controller (SLLPC) and spectral interferometer (SPIDER-DS), which enabled direct and fast measurement of the discrete spectral phase and pulse shaping of the high-power coherent radiation. We also showed that such spectral-phase control produces a train of Fourier-transform-limited pulses (17 fs at full width at half maximum) with an ultrahigh repetition rate of 10.6 THz in the time domain.

### Acknowledgements

The authors acknowledge K. Hakuta for his useful discussions. They also thank N. Yamazaki and N. Sawayama for their help with the experiment. K. R. P. acknowledges the support of a Japanese Government (Monbukagakusho) scholarship. This work was supported partly by the 21st Century COE Program on Coherent Optical Science.